%% file: BHI_Arxiv.tex
\documentclass[letterpaper, 10 pt, conference]{ieeeconf}  %

\IEEEoverridecommandlockouts                              %
\usepackage{graphicx} 
\usepackage{epsfig} 
\usepackage{mathptmx} 
\usepackage{times} 
\usepackage{amsmath} 
\usepackage{amssymb} 
\usepackage{epstopdf}
\usepackage{subfigure}
\usepackage{setspace}
\usepackage{balance}
\usepackage{comment}
\usepackage{blkarray}
\usepackage{tikz}
\usetikzlibrary{intersections}

\numberwithin{subcase}{case}
\renewcommand{\arraystretch}{1.2}
\usetikzlibrary{arrows,shapes,positioning}
\usetikzlibrary{decorations.markings}
\tikzset{->-/.style={decoration={
  markings,
  mark=at position .5 with {\arrow{>}}},postaction={decorate}}}
\usepackage{booktabs}                     
\usepackage{multirow,rotating, array,booktabs}
\renewcommand\footnotemark{}

\title{\LARGE \bf Correlation in Neuronal Calcium Spiking: Quantification based on Empirical Mutual Information Rate}
\author{Sathish Ande$^{*1}$, Srinivas Avasarala$^1$, Ajith Karunarathne$^2$, Lopamudra Giri$^1$,  Soumya Jana$^1$\thanks{$^{1}$Indian Institute of Technology Hyderabad, Telangana, India}
\thanks{$^{2}$ The University of Toledo, Ohio, USA} 
\thanks{$^*$Corresponding author, Email: {\tt\small ee15resch02003@iith.ac.in}
}}

\begin{document} 

\maketitle

\begin{abstract}
\input{section/abstract}

\end{abstract}

\begin{keywords}
 Calcium imaging; Spike train; Empirical probability; Mutual information rate; Correlation.
\end{keywords}

\input{section/introduction}

\input{section/materials}

\input{section/results}

\input{section/conclusion}

\balance
  
\section{ACKNOWLEDGMENT}
We thank Drs. Mennerick and Gautam for providing materials and equipment. Sathish Ande thanks the Ministry of Electronics and Information Technology (MeitY), the Government of India, for fellowship grant under Visvesvaraya PhD Scheme. 

\bibliographystyle{IEEEtran}
\bibliography{ReferencesBib}

\appendix
\section{Proof} 
\input{section/Iproof.tex}
\label{appendix}

\end{document}

%% file: section/abstract.tex
Quantification of neuronal correlations in neuron population helps us to understand neural coding rules. Such quantification could also reveal how neurons encode information in normal and disease conditions like Alzheimer's and Parkinson's. While neurons communicate with each other by transmitting spikes, there would be change in calcium concentration within the neurons inherently. Accordingly, there would be correlations in calcium spike trains and they could have heterogeneous memory structures. In this context, estimation of mutual information rate in calcium spike trains assumes primary significance. However, such estimation is difficult with available methods which would consider longer blocks for convergence without noticing that neuronal information changes in short time windows. Against this backdrop, we propose a faster method which exploits the memory structures in pair of calcium spike trains to quantify mutual information shared between them. Our method has shown superior performance with example Markov processes as well as experimental spike trains. Such mutual information rate analysis could be used to identify signatures of neuronal behaviour in large populations in normal and abnormal conditions.

%% file: section/introduction.tex
\section{INTRODUCTION}
\label{sec:intro}
Neurons are known to encode and process stimulus information in series of action potentials called spike trains. Additionally, such spike trains are shared between neurons for performing high level functions like memory and learning \cite{kennedy2016synaptic}. While doing so, there would be change in calcium ion concentration within each neuron which in turn cause such change in neighbouring neurons in population \cite{berridge1998neuronal,peron2015comprehensive}. Subsequently, quantification of such correlation based calcium spiking between neurons in-terms of mutual information assumes primary significance. Early methods in this direction were based on correlation \cite{cohen2011measuring}. Such methods quantify the linear relation between synchronous spike trains but do not reveal how much information shared between them. Furthermore, spike trains could be non-synchronous. In this context, the methods based upon Lempel-Ziv (LZ) algorithm were suggested to estimate the mutual information between spike trains, which are either synchronous or asynchronous \cite{blanc2008quantifying,szczepanski2011mutual}. However, these methods have slow convergence i.e., they need large length spike trains for estimation, with out identifying the fact that the neuronal information changes in short time windows  \cite{rolls2011neuronal}. In this backdrop, we propose a fast yet accurate empirical method to estimate the mutual information rate by exploiting memory structures in calcium spike trains \cite{sathish2020,sathish2021}. Our method has shown superiority over existing LZ algorithms with example Markov processes and with heterogeneous calcium responses, recorded from hippocampal region of the brain, where memory and learning tasks performed. 
\input{figures/section1/f-CalciumImaging}

\input{figures/section1/f01-flowchart} 
 The rest of this paper is organized as follows. Section~\ref{sec:materials} discusses imaging of neuronal calcium spiking in hippocampal region and deconvolution algorithm for inferring spike trains and introduces the concept of mutual information rate  of stochastic processes, and states LZ algorithms and proposed empirical mutual information rate estimator. Further, Section~\ref{sec:results} describes the results illustrating superiority of the proposed method. Finally, Section~\ref{sec:conclusion} concludes the paper.

%% file: figures/section1/f-CalciumImaging.tex
\begin{figure}[t!]
\centering
\includegraphics[width=0.7\columnwidth]{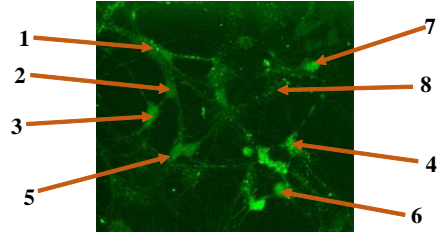}
\label{fig:calciumImaging}
\vspace{-1.2em}
\caption{ Imaging of neuronal calcium spiking: illustrative image of hippocampal neuron population with eight neurons indexed 1-8, as explained in Section \ref{ssec:calcium Imaging}. Scale bar = 20   $\mu$m  \cite{sathish2020}.}
\label{fig:Calcium Imaging}
\end{figure}

%% file: figures/section1/f01-flowchart.tex
\usetikzlibrary{shapes.geometric, arrows}
\tikzstyle{startstop} = [rectangle, rounded corners, minimum width=3cm, minimum height=0.8cm,text centered, draw=black]
\tikzstyle{io} = [trapezium, trapezium left angle=70, trapezium right angle=110, minimum width=3cm, minimum height=1.5cm, text centered, draw=black, fill=blue!30]
\tikzstyle{process} = [rectangle, minimum width=3cm, minimum height=1cm, text centered, draw=black, fill=orange!30]
\tikzstyle{decision} = [diamond, minimum width=3cm, minimum height=1cm, text centered, draw=black, fill=green!30]
\tikzstyle{arrow} = [thick,->,>=stealth]

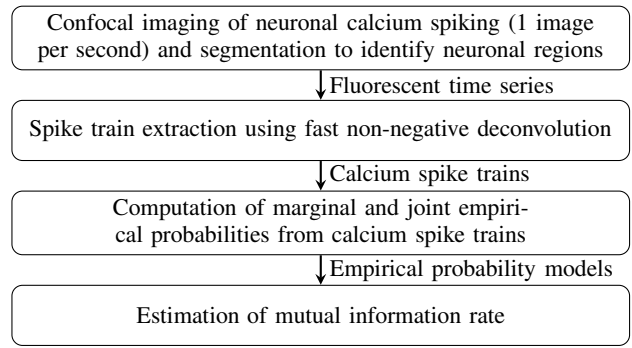
\begin{figure}[t!]
    \centering
   \small
 \begin{tikzpicture}[node distance=35pt]
\node (1) [startstop,text width = 8cm] {Confocal imaging of neuronal calcium spiking (1 image per second) and segmentation to identify neuronal regions};
\node (2) [startstop,below of =1,text width = 8cm] {Spike train extraction using fast non-negative deconvolution};
\node (3) [startstop,below of =2,text width = 8cm] {Computation of marginal and joint empirical probabilities from calcium spike trains};
\node (4) [startstop,below of =3,text width = 8cm] {Estimation of mutual information rate};
\draw [arrow] (1) --node [anchor=west] { Fluorescent time series} (2) ;
\draw [arrow] (2) --node [anchor=west] {Calcium spike trains}(3);
\draw [arrow] (3) -- node [anchor=west] {Empirical probability models}(4);
\end{tikzpicture}
\vspace{-0.5em}
    \caption{Schematic depiction of workflow.}
    \label{fig:workflow}
\end{figure}

%% file: section/materials.tex
\section{MATERIALS AND METHODS}
\label{sec:materials}
The workflow of this paper is shown in Fig. \ref{fig:workflow}, and elaborated in the following.

\subsection{Data Collection and Spike Train Inference}
\label{ssec:calcium Imaging}

The dataset consists of calcium spiking trains from dissociated hippocampal neurons culture from l day postnatal Sprague-Dawley rats. Calcium imaging was performed on 7-th day in vitro (7 DIV) using Leica DMI6000B confocal microscope with Yokogawa CSU-X1 spinning-disk units. Cells were incubated with 1 $\mu$M Fluo-4 (Molecular Probes) dye in 30 min in Hank’s Balanced Salt Solution (HBSS) followed by three washes, each for a duration of 10 min. An excitation and emission wavelength of 488 nm and 510 nm were used for Fluo-4 imaging \cite{giri2014g}. Data acquisition was set at every 1s, but the gap between two images were found to be from 0.8 s to 1 s due to inherent variabilities arising from the instrument. In order to maintain the temperature  at 37$^o$C and 5\% CO$_2$ level, imaging was performed in an incubation chamber attached with the microscope. From the calcium image data (see Fig. \ref{fig:Calcium Imaging} for a illustrative frame), the time course of spatially resolved Fluo-4 fluorescence intensity in neuron populations was obtained using Andor software. 
For the present study, we consider 8 neurons with indices 1--8 from population of 21 neurons for analysis (refer Fig.\ref{fig:fluorescence-2} for time course of calcium responses for aforementioned neurons). We inferred spike train for each neuron from its fluorescent time signal using a deconvolution algorithm and those were binarized by adaptive threshold  \cite{sathish2020,vogelstein2010fast}. Such binary spike trains were used for quantification analysis.
\input{figures/section2/f02-hippocampus-fluorescent-time-series}


\subsection{Mathematical Preliminaries} 
\label{sec:math}

At this point, we introduce the concept of mutual information rate, and state existing mutual information rate estimators and proposed estimator. 
\subsubsection{Entropy rate} Entropy,  a measure of uncertainty in random variable $X$ with probability mass function $P_X$ defined on alphabet $\mathcal{X}$ is defined by  $H(X)=-\sum_{x\in \mathcal{X}} p_X(x)\log p_X(x)$. The entropy per symbol, entropy rate, of sequence of random variables $X_1,X_2,\hdots,X_n,\hdots$ is defined as 
\begin{align} \label{eq: entropyRate}
\bar{H}(X)=\lim _{n\rightarrow \infty} \frac{1}{n}H(X^n),
\end{align}
$X^n=(X_1,X_2\hdots,X_n)$.Thus 
$\bar{H}(X)$ measures the average asymptotic uncertainty per symbol \cite{cover2012elements}.  
\subsubsection{Mutual information rate}
\label{ssec:Entropyrate}
 Mutual information $I(X;Y)$, a measure of mutual dependency between two random variables $X$, $Y$ with probability mass functions $P_X$, $P_Y$ defined on alphabets ${\mathcal{X}}$, ${\mathcal{Y}}$ respectively is defined as  
 \begin{align} 
I(X;Y)= \sum_{x\in \mathcal{X}} \sum_{y\in \mathcal{Y}} P_{XY}(x,y)\log \frac{P_{XY}(x,y)}{P_X P_Y}
\end{align}
where $P_{XY}(x,y)$ is joint probability mass function of $X$, $Y$. It can be written in-terms of marginal and joint entropy's of $X$, $Y$ as $ I(X;Y) = H(X) + H(Y) - H(X,Y)$.

 Further, the mutual information rate  $\bar{I}(X;Y)$ of a sequence of random variables $X_1,X_2,\hdots,X_n,\hdots$ and $Y_1,Y_2,\hdots,Y_n,\hdots$ is defined as
\begin{align} \label{eq: mutualRate}
\bar{I}(X;Y)=\lim _{n\rightarrow \infty} \frac{1}{n}I(X^n;Y^n),
\end{align}
where $X^n= X_1^n=(X_1,X_2\hdots,X_n)$, $Y^n=Y_1^n=(Y_1,Y_2\hdots,Y_n)$. Thus 
$\bar{I}(X;Y)$ measures the asymptotic reduction in uncertainty per symbol \cite{cover2012elements}. It can be written in-terms of marginal and joint entropy rates of $X^n$, $Y^n$ as follows.
\begin{align}
\bar{I}(X;Y)&=\lim _{n\rightarrow \infty} \frac{1}{n}I(X^n;Y^n) \nonumber\\
&= \lim _{n\rightarrow \infty}(\frac{1}{n}H(X^n)+ \frac{1}{n} H(Y^n)- \frac{1}{n}H(X^n,Y^n))
\label{eq:muDef}
\end{align}

\subsubsection{
Lempel-Ziv estimators of mutual information rate}
\label{sssec:LZ}
Versions of Lempel-Ziv (LZ) algorithms have been proposed for data compression, which find the complexity $K_{LZ}$ of the sequence $X^n$, a measure of entropy rate. \\
\underline{LZ-78 algorithm}:
LZ-78 is dictionary based scheme that converts a given sequence into unforeseen sub-blocks to the dictionary.
\input{figures/section2/LZtable}
\input{figures/section2/t01-empirical-model}
Denoting by $b(X^n)$ the number of sub-blocks of a sequence $X^n$ present in the data, the LZ-78 complexity, the average number of bits per symbol, is defined as \cite{cover2012elements}
\begin{align}
K_{LZ78}(X^{n})= \frac{b(X^{n})\log(b(X^{n}))}{n}.
\label{eq:LZ78}
\end{align}
So, the corresponding entropy rate estimate is $\bar H_{LZ78}(X) =\lim_{n\rightarrow \infty} K_{LZ78}(X^{n})$, where one practically uses large $n$ in place of the limit. The mutual LZ-78 complexity $K_{LZ78}(X^n,Y^n)$ between sequences $X^n$,$Y^n$  is given as \cite{blanc2008quantifying}
\begin{align}
K_{LZ78}(X^{n},Y^n)= \frac{b(X^{n},Y^n)\log(b(X^{n},Y^n))}{n}.
\label{eq:MLZ78}
\end{align}
where $b(X^{n},Y^n)$ is number of sub-blocks in joint dictionary of $X^{n},Y^n$. The mutual information rate is given as
\begin{align}
\bar I_{LZ78}(X;Y)= \lim_{n\rightarrow \infty} (K_{LZ78}(X^{n}) +  K_{LZ78}(Y^{n})- K_{LZ78}(X^{n},Y^n)) 
\label{eq:ILZ78}
\end{align}
\\
\underline{Sliding Window Lempel Ziv (SWLZ) algorithm}:
SWLZ converts the given sequence into sub-blocks such that  new block is seen at each position $i$ is not appear in previous $i-1$ symbols by sliding a window of length, $L_i$. 
The SWLZ complexity is given by \cite{sathish2020}
\begin{align}
K_{SWLZ}(X^n) =  \sum_{i=1}^{n} \frac{n\log n}{L_i(X^n)},
\label{eq:LZ77}
\end{align}
So, the corresponding entropy rate estimate is $\bar H_{SWLZ}(X) =\lim_{n\rightarrow \infty} K_{SWLZ}(X^{n})$. The mutual SWLZ complexity $K_{SWLZ}(X^n,Y^n)$ is given by \cite{sathish2020}
\begin{align}
K_{SWLZ}(X^n,Y^n) =  \sum_{i=1}^{n} \frac{2n\log n}{L_i(X^n,Y^n)},
\label{eq:MLZ77}
\end{align}
where $L_i(X^{n},Y^n)$ is length of sub-block at instant $i$ in joint dictionary of $X^{n},Y^n$.
The mutual information rate is given as
\begin{align}
\bar I_{SWLZ}(X;Y)= \lim_{n\rightarrow \infty} (K_{SWLZ}(X^{n}) +  K_{SWLZ}(Y^{n})- K_{SWLZ}(X^{n},Y^n)) 
\label{eq:ILZ77}
\end{align}
\\
\underline{LZ-76 algorithm}:
LZ-76 provides another version of the dictionary-based scheme LZ-78. The LZ-76 complexity is given by  \cite{amigo2004estimating}
\begin{align}
K_{LZ76}(X^n) = \frac{b(X^n)}{n} \log n,
\label{eq:LZ76}
\end{align}
where $b(X^n)$ represents total number of sub-blocks. So, the corresponding entropy rate estimate is $\bar H_{LZ76}(X)=\lim_{n\rightarrow \infty} K_{LZ76}(X^{n})$.
The mutual LZ-76 complexity $K_{LZ78}(X^n,Y^n)$ is given as 
\begin{align}
K_{LZ76}(X^n,Y^n) = \frac{b(X^n,Y^n)}{n} \log n, 
\label{eq:MLZ76}
\end{align}
here $b(X^{n},Y^n)$ is number of sub-blocks in joint dictionary of $X^{n},Y^n$.
The mutual information rate is given as
\begin{align}
\bar I_{LZ76}(X;Y)= \lim_{n\rightarrow \infty} (K_{LZ76}(X^{n}) +  K_{LZ76}(Y^{n})- K_{LZ76}(X^{n},Y^n)) 
\label{eq:ILZ76}
\end{align}
\\
\underline{Illustration}: Parsing by the aforementioned algorithms are illustrated on an example strings $X^n=        `010011001011101$', $Y^n=`100110010100100$' and the joint string $(X^n,Y^n)=`011000011110000110011010110010$' in Table \ref{table:LZ787776}.
\input{figures/section3/f04-rasterplot-example2}

\input{figures/section3/f05-example2-comparison}
\subsection{Proposed Estimator based on Empirical Probabilities}
\label{sssec:proposed}
 we propose an estimator $\bar I_{EP}(X;Y)$ which models the neuronal spiking process as Markov process up to a prescribed order $k$. Further, the conditional probabilities of different contextual symbols up to a prefixed order $k$ are calculated from sequences empirically. For instance, the marginal and joint empirical probability models for the sequences $X^n= `010011001011101$', $Y^n=`100110010100100$' and the joint string $(X^n,Y^n)=`011000011110000110011010110010$' are shown in Tables \ref {table:EmX}, \ref{table:EmY}, \ref{table:EmXY} respectively. Here, the $k=1$. \\ The mutual information estimator $\bar I(X;Y)$ for sequences $X^n= X_1^n=(X_1,X_2\hdots,X_n)$ and $Y^n=Y_1^n=(Y_1,Y_2\hdots,Y_n)$ is as follows.
 \begin{equation}
\begin{aligned}
\bar{I}_{EP}(X;Y) &= \bar{H}_{EP}(X) + \bar{H}_{EP}(X) -\bar{H}_{EP}(X,Y)\\
&= \frac{1}{n}H(X_1^n)+ \frac{1}{n}H(Y_1^n)-\frac{1}{n}H(X_1^n ,Y_1^n)\\
\end{aligned}
\label{eq:EMI} 
\end{equation}
where 
\begin{equation}
\begin{aligned}
 \frac{1}{n}H(X_1^n,Y_1^n)&= \frac{1}{n}(H(X_1,Y_1 )+...+H(X_{k+1},Y_{k+1}\mathbin{\vert}X_1^k,Y_1^k)... \\
&  ...+ H(X_{n},Y_n\mathbin{\vert}X_1^{n-1},Y_1^{n-1} )  \quad \quad \text{by chain rule} \\
&= \frac{1}{n}(H(X_1 )+H({X_2}\mathbin{\vert}{X_1} )+.....\\  & + (n-k) H(X_{k+1},Y_{k+1}\mathbin{\vert}X_1^k,Y_1^k))\quad \text{for order $k$}\\ 
& \approx H(X_{k+1},Y_{k+1}\mathbin{\vert} X_1^k,Y_1^k) \quad \text{if $n>>k$},
\end{aligned}
\label{eq:EmJointE} 
\end{equation}
where ($X_1^k,Y_1^k$) provides the context of order $k$ (i.e., $k$ previous symbols) for $X_{k+1},Y_{k+1}$. In the preceding, $H(U\mathbin{\vert}V)$ denotes the conditional entropy of $U$ given $V$.
similarly, the entropy rate estimators for $X^n$ and $Y^n$  are as follows \cite{sathish2020}.
\begin{equation}
\begin{aligned}
 \frac{1}{n}H(X_1^n) \approx H(X_{k+1}\mathbin{\vert} X_1^k) \quad \text{if $n>>k$},\\
 \frac{1}{n}H(Y_1^n) \approx H(Y_{k+1}\mathbin{\vert} Y_1^k) \quad \text{if $n>>k$},
\end{aligned}
\label{eq:EmJointE} 
\end{equation}
Hence, the mutual information estimator $\bar I(X;Y)$ is

\begin{equation}
\begin{aligned}
\bar{I}_{EP}(X;Y) \approx H(X_{k+1}\mathbin{\vert} X_1^k)+H(Y_{k+1}\mathbin{\vert} Y_1^k)- \\ H(X_{k+1},Y_{k+1}\mathbin{\vert} X_1^k,Y_1^k)
\label{eq:EmMI}
\end{aligned}
\end{equation}

%% file: figures/section2/f02-hippocampus-fluorescent-time-series
\begin{figure}[t!]
\centering
\includegraphics[width=\columnwidth]{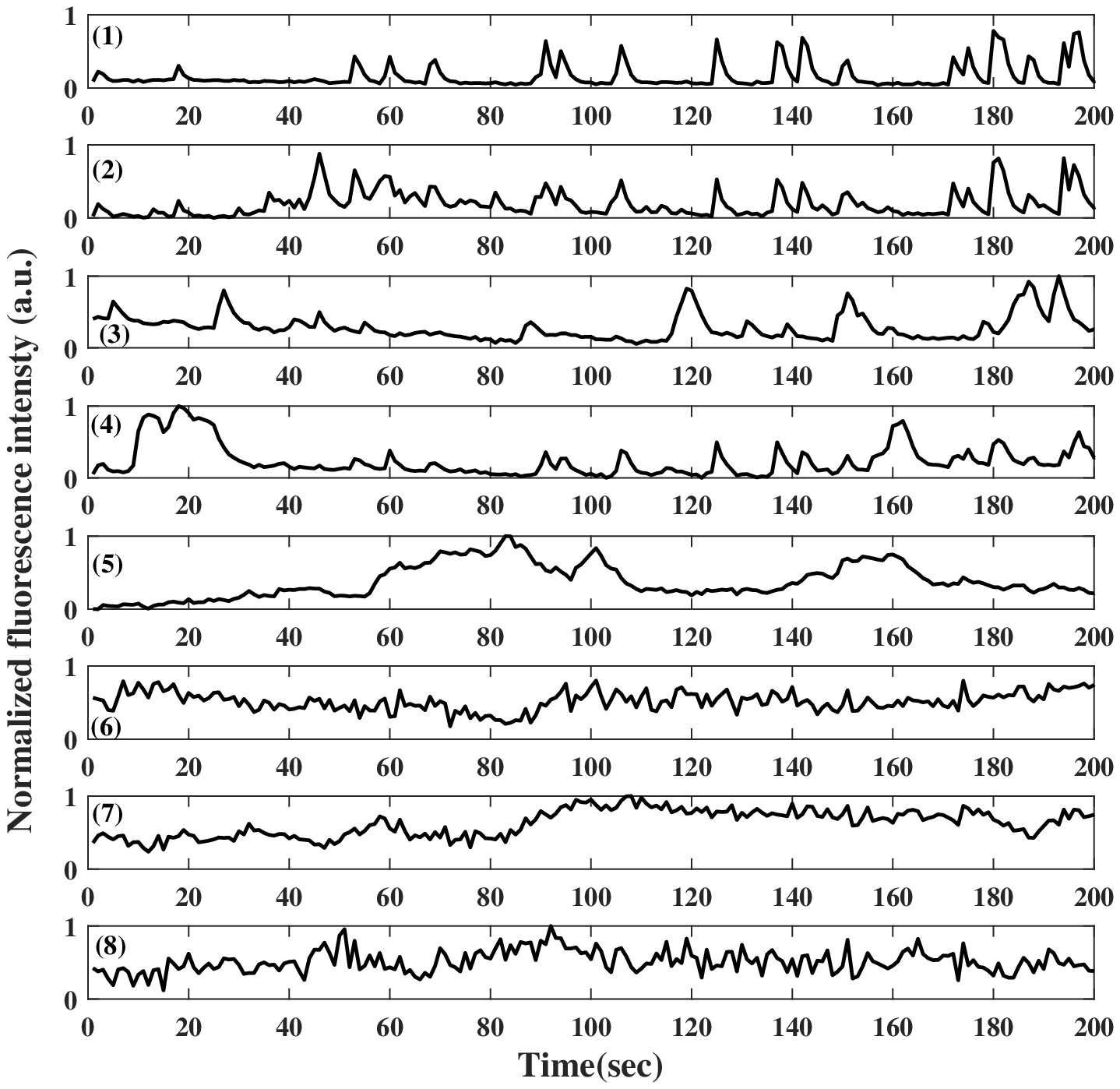}
 \vspace{-3.4em}
\caption{Time course of normalized Fluo-4 intensity for eight neurons indexed 1-8 in Fig: \ref{fig:calciumImaging}, as explained in Section \ref{ssec:calcium Imaging}.}
\label{fig:fluorescence-2}
\end{figure}

%% file: figures/section2/LZtable.tex
\begin{table*}[t]
\small
\begin{tabular}{|c|c|c|c|}
\hline
Algorithm & Dictionary for $X^n$                                                                                          & Dictionary for $Y^n$                                                                                               & Dictionary for $(X^n, Y^n)$                                                                                                           \\ \hline
LZ-78     & 0|1|00|11|001|01|110|1                                                                                  & 1|0|01|10|010|100|100                                                                                        & \begin{tabular}[c]{@{}c@{}}01|10|00|0111|1000|0110|\\ 011010|11|0010\end{tabular}                                           \\ \hline
SWLZ      & \begin{tabular}[c]{@{}c@{}}0|1|00|011|11|10010|\\ 0010|0101|1011|\\ 0111|111|1101|101|01|1\end{tabular} & \begin{tabular}[c]{@{}c@{}}1|0|01|11|10010|0010|\\ 010|101|\\ 010|100100|00100|0100|\\ 100|00|0\end{tabular} & \begin{tabular}[c]{@{}c@{}}01|10|00|0111|11|10000110\\ |000110\\ |011001|1001|011010|1010|\\ 1011|1100|0010|10\end{tabular} \\ \hline
LZ-76     & 0|1|00|11|0010|111|01                                                                                   & 1|0|01|10010|100100                                                                                          & \begin{tabular}[c]{@{}c@{}}01|10|00|0111|10000110|\\ 011010|1100|10\end{tabular}                                           \\ \hline
\end{tabular}
\caption{LZ-78,  SWLZ and LZ-76 algorithms:  Parsing of example sequences $X^n=`010011001011101$', $Y^n=`100110010100100$'.}
\label{table:LZ787776}
\end{table*}

%% file: figures/section2/t01-empirical-model.tex
\begin{table}[t!]
\centering
\caption{Empirical probability model of $X^n$}
\small
\renewcommand*{\arraystretch}{1.1} 
\resizebox{\columnwidth}{!}{
\begin{tabular}{ccccccc} \toprule
\multicolumn{3}{c}{k=0 (no context)}                                          & \multicolumn{4}{c}{k=1}                                                                 \\ \bottomrule\\
Symbol & Count & \begin{tabular}[c]{@{}c@{}}Relative\\ frequency\end{tabular} & Context & Symbol & Count & \begin{tabular}[c]{@{}c@{}}Relative\\ frequency\end{tabular} \\\bottomrule\\
0      & 7     & 7/15                                                         & 0       & 0      & 2     & 2/7                                                          \\
1      & 8     & 8/15                                                         & 0       & 1      & 5     & 5/7                                                          \\
       &       &                                                              & 1       & 0      & 4     & 4/7                                                          \\
       &       &                                                              & 1       & 1      & 3     & 3/7   \\\bottomrule                                                     
\end{tabular}}
\label{table:EmX}
\end{table}

\begin{table}[t!]
\centering
\caption{Empirical probability model of $Y^n$}
\small
\renewcommand*{\arraystretch}{1.1} 
\resizebox{\columnwidth}{!}{
\begin{tabular}{ccccccc} \toprule
\multicolumn{3}{c}{k=0 (no context)}                                          & \multicolumn{4}{c}{k=1}                                                                 \\ \bottomrule\\
Symbol & Count & \begin{tabular}[c]{@{}c@{}}Relative\\ frequency\end{tabular} & Context & Symbol & Count & \begin{tabular}[c]{@{}c@{}}Relative\\ frequency\end{tabular} \\ \bottomrule\\
0      & 9     & 9/15                                                         & 0       & 0      & 4     & 4/8                                                          \\
1      & 6     & 6/15                                                         & 0       & 1      & 4     & 4/8                                                          \\
       &       &                                                              & 1       & 0      & 5     & 5/6                                                          \\
       &       &                                                              & 1       & 1      & 1     & 1/6   \\\bottomrule                                                      
\end{tabular}}
\label{table:EmY}
\end{table}
\begin{table}[t!]
\centering
\caption{Empirical probability model of ($X^n,Y^n$)}
\small
\renewcommand*{\arraystretch}{1.1} 
\resizebox{\columnwidth}{!}{
\begin{tabular}{ccccccc}
\toprule
\multicolumn{3}{c}{k=0 (no context)}                                          & \multicolumn{4}{c}{k=1}                                                                 \\
\bottomrule \\
Symbol & Count & \begin{tabular}[c]{@{}c@{}}Relative\\ frequency\end{tabular} & Context & Symbol & Count & \begin{tabular}[c]{@{}c@{}}Relative\\ frequency\end{tabular} \\
\bottomrule\\
00     & 3     & 3/14                                                         & 00      & 00     & 0     & 0                                                            \\
01     & 4     & 4/14                                                         & 00      & 01     & 2     & 2/3                                                          \\
10     & 5     & 5/14                                                         & 00      & 10     & 1     & 1/3                                                          \\
11     & 2     & 2/14                                                         & 00      & 11     & 0     & 0                                                            \\
       &       &                                                              & 01      & 00     & 0     & 0                                                            \\
       &       &                                                              & 01      & 01     & 0     & 0                                                            \\
       &       &                                                              & 01      & 10     & 3     & 3/4                                                          \\
       &       &                                                              & 01      & 11     & 1     & 1/4                                                          \\
       &       &                                                              & 10      & 00     & 2     & 2/5                                                          \\
       &       &                                                              & 10      & 01     & 1     & 1/5                                                          \\
       &       &                                                              & 10      & 10     & 1     & 1/5                                                          \\
       &       &                                                              & 10      & 11     & 1     & 1/5                                                          \\
       &       &                                                              & 11      & 00     & 1     & 1/2                                                          \\
       &       &                                                              & 11      & 01     & 0     & 0                                                            \\
       &       &                                                              & 11      & 10     & 1     & 1/2                                                          \\
       &       &                                                              & 11      & 11     & 0     & 0         \\\bottomrule                                                  
\end{tabular}}
 \label{table:EmXY}
\end{table}

%% file: figures/section3/f04-rasterplot-example2.tex
\begin{figure}[ht!]
\centering
\includegraphics[width=\columnwidth]{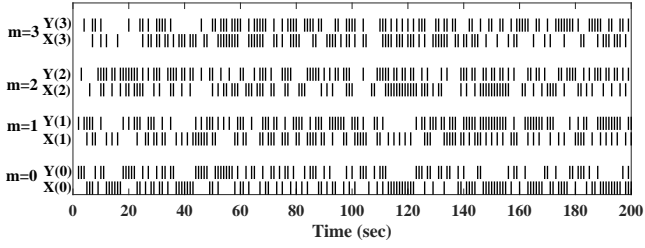}
\vspace{-2.7em}
\caption{Raster plots of $m$-th ($m=0,1,2,3$) order sources given in the example in Section \ref{ssec:simulation}.}
\label{fig:Rasterplot_ex2} 
\end{figure}

%% file: figures/section3/f05-example2-comparison.tex
 \begin{figure*}[t!]
\centering
\begin{tabular}{cccc}
\multicolumn{4}{c}{
\includegraphics[width=15cm]{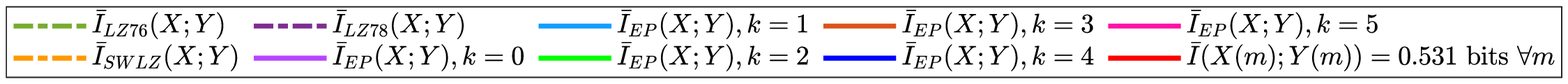}
}
\vspace{-0.9em}
\\ 
\hspace*{-4em}

\includegraphics[width=4.8cm]{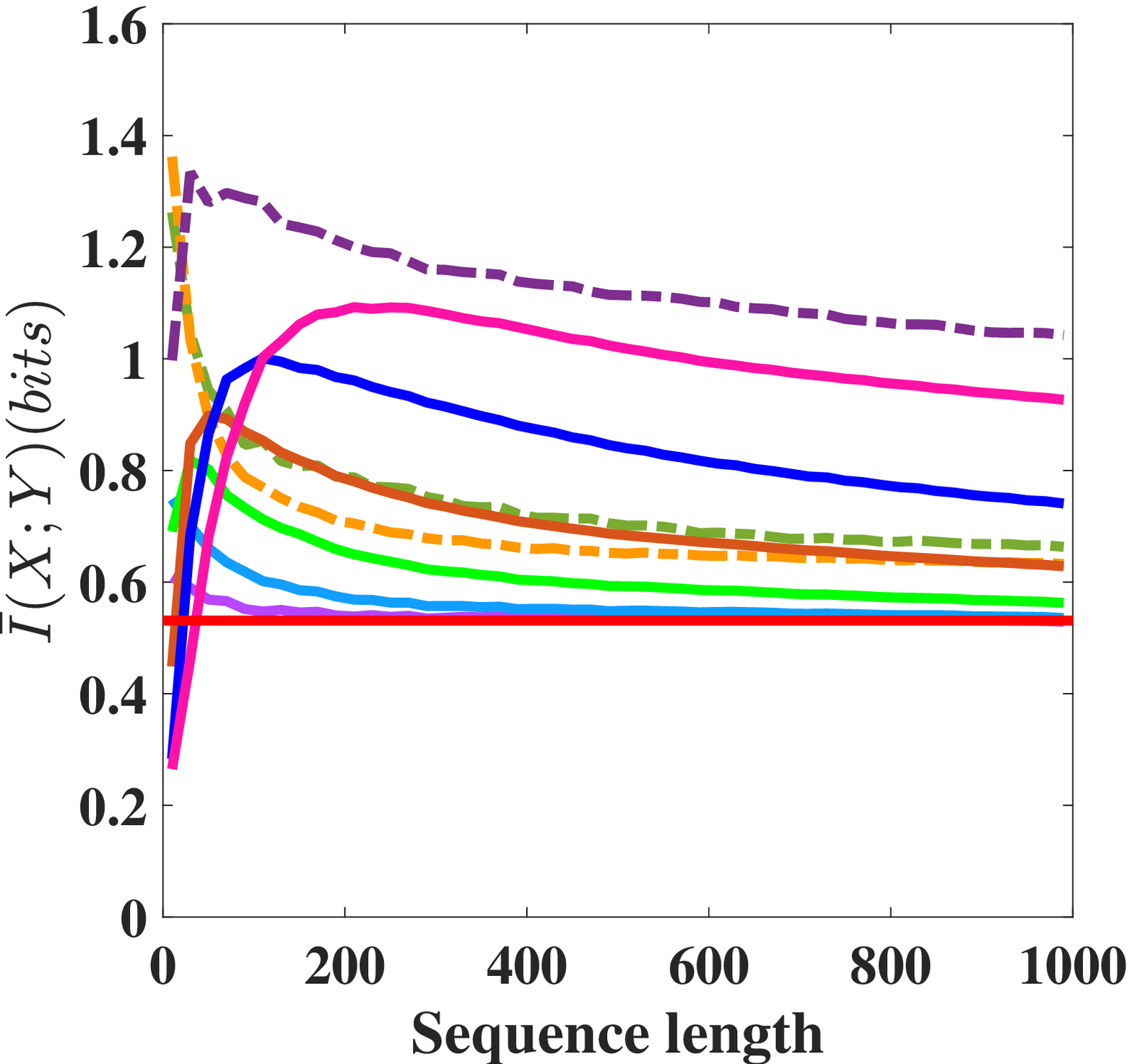}
\hspace*{-3.4em}
&

\includegraphics[width=4.8cm]{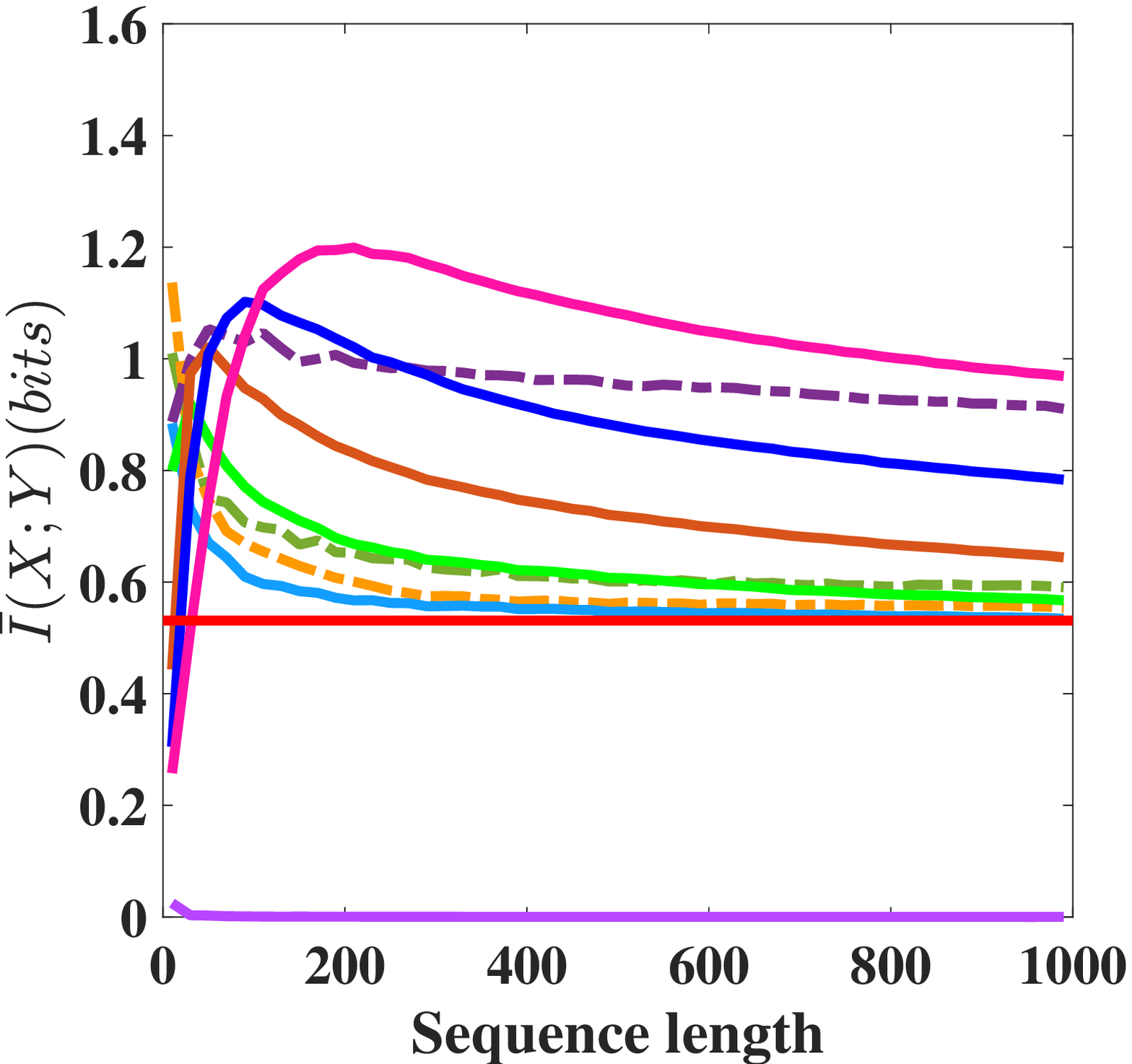}
\hspace*{-2.25em}
&
\includegraphics[width=4.8cm]{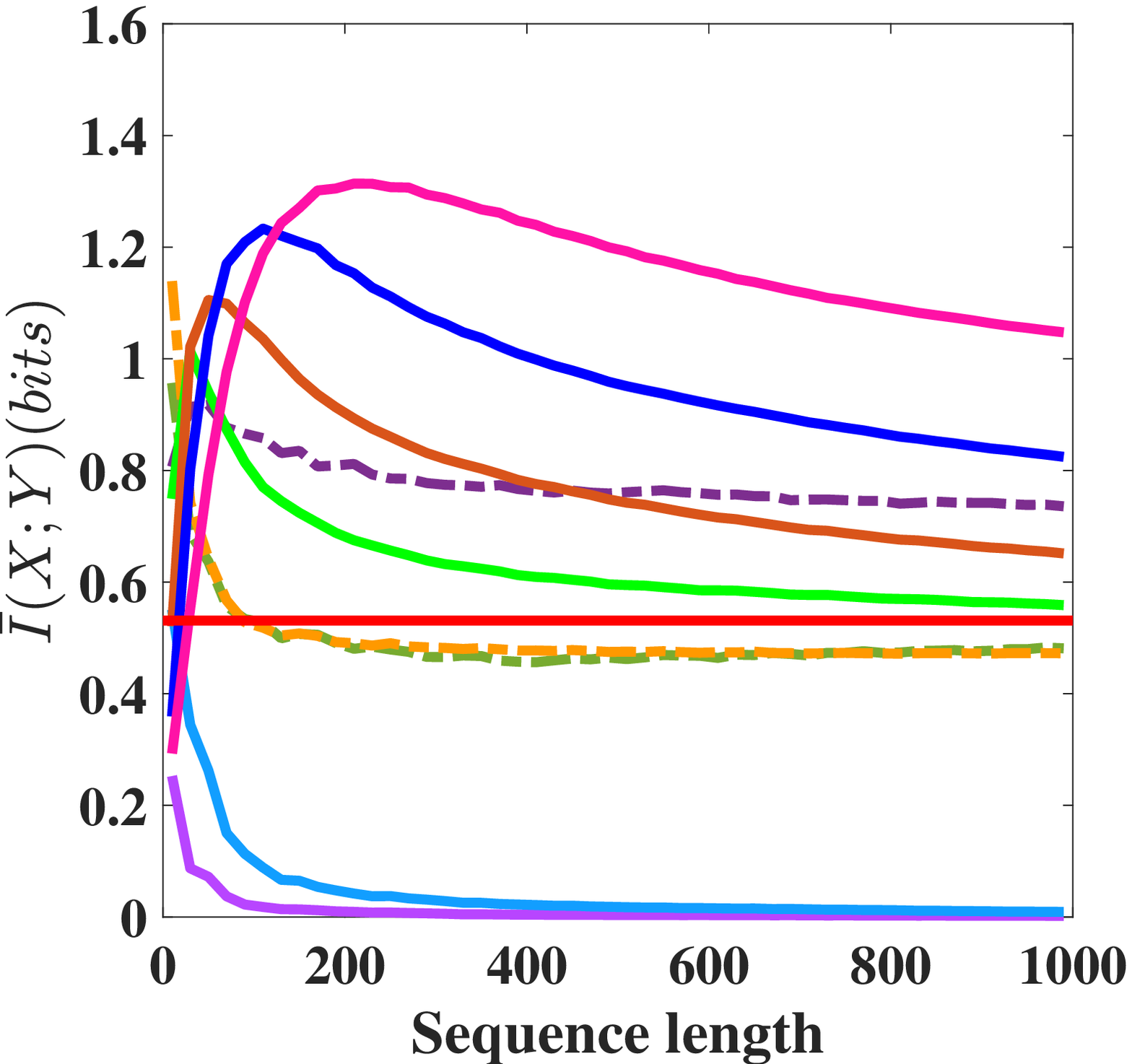}
\hspace*{-2.25em}
&
\includegraphics[width=4.8cm]{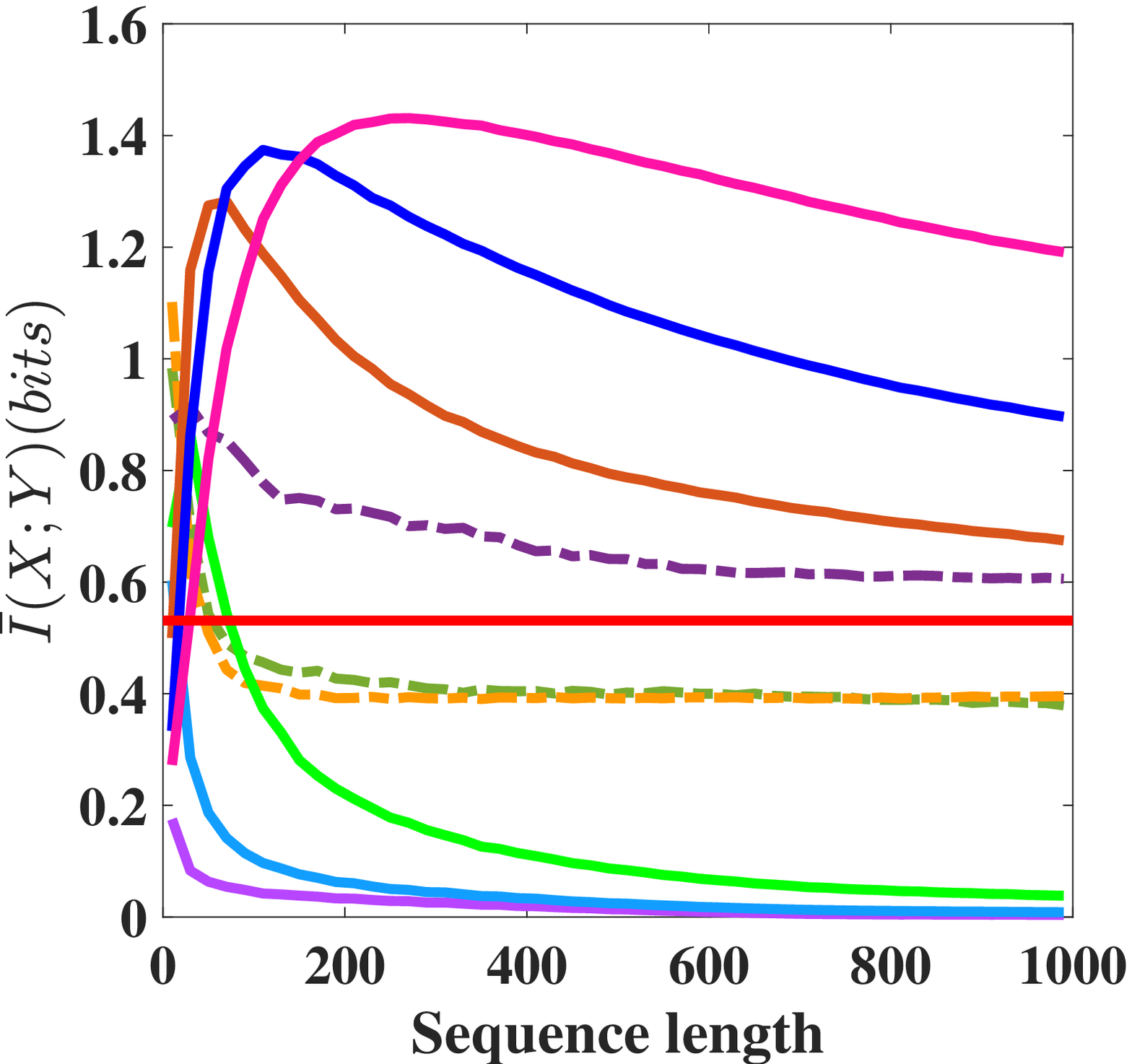}
\vspace*{-0.5em}
\\
\footnotesize{(a)  $\bar{I}(X(0);Y(0))= 0.531$ bits.}
&
\footnotesize{(b) $\bar{I}(X(1);Y(1))= 0.531$ bits.}
&
\footnotesize{(c) $\bar{I}(X(2);Y(2))= 0.531$ bits.} 
&
\footnotesize{(d) $\bar{I}(X(3);Y(3))= 0.531$ bits.}
\end{tabular}
\caption{Comparison of the proposed mutual information rate estimate $\bar{I}_{EP}(X;Y)$ with estimates based on LZ versions $\bar{I}_{LZ78}(X;Y)$, $\bar{I}_{SWLZ}(X;Y)$, $\bar{I}_{LZ76}(X;Y)$,  with context lengths $k=0,1,2,3,4,5$ for $m$-th order Markov sources: (a)-(d) for $m=0,1,2,3$ respectively.}
\label{fig:simulation2_m1234}
\end{figure*}


%% file: section/results.tex
\section{RESULTS}
\label{sec:results}
 We begin by comparing our method with existing LZ algorithms with an example Markov process. Subsequently, we extend such comparison to neuronal calcium spike trains .
\subsection{Results for Example Markov Process}
\label{ssec:simulation}
 We compared estimates by generating four-state (00, 01, 10 and 11) Markov processes ($X(m), Y(m)$) (the number in parenthesis indicating the order) of order $m$ with mutual information rate $\bar I(X(m);Y(m))$. $X(m)$, $Y(m)$ were generated by using following equations
\begin{align}
 X_i = X_{i-1} \oplus X_{i-2} \oplus ..........\oplus X_{i-m+1} \oplus X_{i-m} \oplus u_i
\label{eq:markov1}
\end{align}
\begin{align}
 Y_i = Y_{i-1} \oplus Y_{i-2} \oplus ..........\oplus Y_{i-m+1} \oplus Y_{i-m} \oplus v_i
\label{eq:markov2}
\end{align}
where ($u_{i},v_i$) is an joint source with distribution $Pr(u,v)$ in (\ref{EqUV}) and the  `$\oplus$'  indicates modulo-2 addition.

\begin{equation}
Pr(u,v) =\begin{blockarray}{ccc}
& 0 & 1  \\
\begin{block}{c(cc)}
 0 & 0.25-p & 0.25+p   \\
1 &  0.25+p & 0.25-p   \\
\end{block}
\end{blockarray}
\label{EqUV}
\end{equation}
Hence, \\ $P(u=0) = P(u=1) =0.5$ and $P(v=0)=P(v=1)=0.5$. 
 \input{figures/section2/f03-deconvolution} 
\input{figures/section3/t02-std-estimates}
\input{figures/section3/f06-stationary}

\input{figures/section3/f-rasterplot-neurons}

\input{figures/section3/f07-temporal-variation-estimates} 
The designed process has property that 
\begin{equation}
\begin{aligned}
\bar{I}(X;Y) &= I(u;v) \\
&= 2(1+(a_1) \log_2 (a_1) +(a_2) \log_2 (a_2) ) \quad \forall n, m
\end{aligned}
\label{Eq:MIuv}
\end{equation}
where $a_1 = 0.25 - p$ and  $a_1 = 0.25 + p$.
 The mutual information rate between $X(m)$, $Y(m)$ is equal to  mutual information rate between binary sources $u_{i}$ and $v_i$. Refer appendix for proof of (\ref{Eq:MIuv}). 

Mutual information rate of these processes $\bar I(X(m),Y(m))=$0.531 bits $\forall m$ for chosen $p=$0.2 using (\ref{Eq:IMarkov}). The raster plot for a realization of each of the above processes for $m=0,1,2,3$ generated by using (\ref{eq:markov1}), (\ref{eq:markov2}) is shown in Fig. \ref{fig:Rasterplot_ex2}.

We next demonstrate the efficacy of the proposed estimator. Specifically, we calculated $\bar I_{EP}(X(m);Y(m))$ using (\ref{eq:EmMI}), for each of the context lengths $k=0,1,2,3$ and each of the generated processes $(X(0),Y(0))$, $(X(1),Y(1))$, $(X(2),Y(2))$, $(X(3),Y(3))$ (corresponding to $m=0,1,2,3$, respectively), and depict those against the window (sequence) length in Figures \ref{fig:simulation2_m1234}(a)-(d), respectively. In the respective figures, we compared the proposed method against various LZ estimates, $\bar I_{LZ78}(X;Y)$, $\bar I_{SWLZ}(X;Y)$, and $\bar I_{LZ76}(X;Y)$ (see (\ref{eq:LZ78}), (\ref{eq:LZ77}) and (\ref{eq:LZ76}) respectively). Our method
tends to converge to $\bar I(X(\min(m,k);Y(\min(m,k))$, for each pair $(k,m)$, as designed.
In particular, we underestimated mutual information rate when the order was underestimated, i.e., $k < m$. For instance, in Figure $5(c)$, while $\bar I(X(2);Y(2))
=$0.531 bits for $m=2$, we underestimate $\bar I_{EP}(X(2);Y(2))\approx$ 0 bits for $k=0$ and $k=1$. 
However, when $k\ge m$, the proposed estimator was accurate, and the convergence to $\bar I(X(m);Y(m))$ was faster than the each of the reference algorithms for lower $k$. 
For instance, in Figure $5(a)$, $\bar I_{EP}(X(0);Y(0))$ ($m=0$)  approximately converges to $\bar I(X(0);Y(0))=$ 0.531 bits for each of $k=0,1,2$, faster than reference estimates.This case would also arise for higher $k$ in experimental spike trains explained in section \ref{ssec:experimental}.  Among the rest, the nearest competitor is $\bar I_{SWLZ}(X;Y)$, which shows the fastest convergence among LZ variants, and will be used for comparison in subsequent analysis.

 \subsection{Experimental Results}
\label{ssec:experimental}
 For experimental analysis, we chose eight neurons, indexed 1, 2, 3, 4, 5, 6, 7, 8 (see Fig.\ref{fig:Calcium Imaging}) whose calcium responses are shown in Fig. \ref{fig:fluorescence-2}. As explained earlier, We first applied the deconvolution method to infer spike trains, and computed the spiking rates and then spiking threshold $P_{th}=0.36$  to remove spurious spikes. In Fig. \ref{fig:spiketrain}, we showed the inferred spike train for neuron 1, and analogous binary spike sequences were used for further analysis.

\subsubsection{Optimization of block length}
\label{sssec:stationarity}
We studied the behavior of the proposed estimator $\bar I_{EP}(X;Y)$ for neurons 1, 2 for different lengths of time windows, and compared it with $\bar I_{SWLZ}(X;Y)$, as stated earlier. For that purpose, the midpoint of all time windows of length 50, 100, 150, 200 was fixed at 115 and shifted by 5 every time up to 160. Here, we varied the context length within $k=0,1,2$. Subsequently, we estimated mutual information rates in the aforementioned time windows for different $k$, whose standard deviations  are furnished in Table~\ref{table:stdestimates}. We observed that mutual information rate estimates had smallest standard deviation for windows of length 200 for each $k$. Such behaviour was clearly corroborated in Figure \ref{fig:estimates_windows_k012}. Note that $\bar I_{SWLZ}(X;Y)$ exhibited relatively less deviation in windows of length 200 but it also always overestimated the mutual information rate, and proved unsuitable for present comparison. Thus, the window length of 200 is most well suited for mutual information rate estimation. However, such analysis should be carried out with more neurons for more comprehensive conclusions.  

  \subsubsection{Temporal variation in mutual information rate estimates}
\label{sssec:Temporal}
We next turn to studying temporal variation in mutual information content. To this end, we again considered with aforementioned neurons $1,2,3,4,5,6,7,8$, and the spike sequences are shown in raster plot in Fig. \ref{fig:Rasterplot_neurons}. Specifically, we  plotted temporal variation of mean estimated mutual information rates of sequence pairs 1-2, 3-4, 5-6, 7-8 using  $\bar I_{EP}(X;Y)$ and $\bar I_{SWLZ}(X;Y)$ in Fig. \ref{fig:TempVariation}. Here, the context length $k$ was varied from 0 to 6. The mutual information rate estimates, $\bar I_{EP}(X;Y)$ converges to a value between sequence lengths, 150 and 200 unlike $\bar I_{SWLZ}(X;Y)$. For neuron 5-6, the mutual information rate estimates, $\bar I_{EP}(X;Y)$  were almost same across time for the orders $k=0,1,...,5$ unlike for $k=6$, for which slight overestimation is observed (Fig. \ref{fig:nd3}). Further, there was some gap in estimates, $\bar I_{EP}(X;Y)$, $k=0$, and $\bar H_{EP}(X)$, $k=1,...,6$ for neurons 1-2 and 3-4 as shown in Figures \ref{fig:nd1} and \ref{fig:nd2}.  
This suggested that there could be process memory of order $k= 1$ for neurons 1-2 and 3-4 unlike neurons 5-6, for which order $k=0$ appears enough. However, the temporal behaviour of neurons 7-8 was completely different from the aforementioned neurons as shown in Figure \ref{fig:nd4}. The proposed mutual information rate estimates $\bar I_{EP}(X;Y)$ were different and underestimated for each $k=0,1,2$. Further, such estimates were equal to  $\bar I_{SWLZ}(X;Y)$ for $k=4,5,6$. This suggested that there could be complex process memory structure for this type of neurons. In other words, we clearly observed heterogeneity in mutual information rate estimates, considering only eight neurons. We believe that the proposed estimator can provide a useful tool for studying heterogeneity in process memory in large neuron populations.

%% file: figures/section2/f03-deconvolution.tex
\begin{figure}[t!]
\centering  \includegraphics[width=\columnwidth]{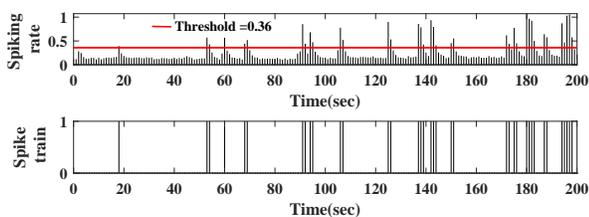} \vspace{-2.5em}
  \caption{Spike train inferred for neuron 1 from fluorescence intensity data using the aforesaid deconvolution method \cite{vogelstein2010fast} (see top panel in Figure \ref{fig:fluorescence-2}).} 
\label{fig:spiketrain}
\end{figure}

%% file: figures/section3/t02-std-estimates.tex
\textit{\begin{table}[t!]
\centering
\small
\caption{Standard devation of estimated mutual information rates for time windows of different lengths for the neurons indexed 1, 2}
\begin{tabular}{|c|c|c|c|c|}
\hline
\multicolumn{5}{|c|}{Standard deviation,       $\sigma_{I}$}                                                                                     \\ \hline
\multirow{2}{*}{\begin{tabular}[c]{@{}c@{}}Block \\ length\end{tabular}} & \multirow{2}{*}{$\bar I_{SWLZ}(X;Y)$} & \multicolumn{3}{c|}{$\bar I_{EP}(X;Y)$} \\ \cline{3-5} 
                                                                         &                       & k=0     & k=1    & k=2    \\ \hline
50                                                                       & 0.151                 & 0.040   & 0.095  & 0.061  \\ \hline
100                                                                      & 0.089                 & 0.037   & 0.032  & 0.035  \\ \hline
150                                                                      & 0.038                 & 0.022   & 0.030  & 0.024  \\ \hline
200                                                                      & 0.029                 & 0.021   & 0.012  & 0.014  \\ \hline
\end{tabular}
\label{table:stdestimates}
\end{table}}

%% file: figures/section3/f06-stationary.tex
\begin{figure}[t!]
  \centering
\subfigure{
 \includegraphics[width=0.8\columnwidth]{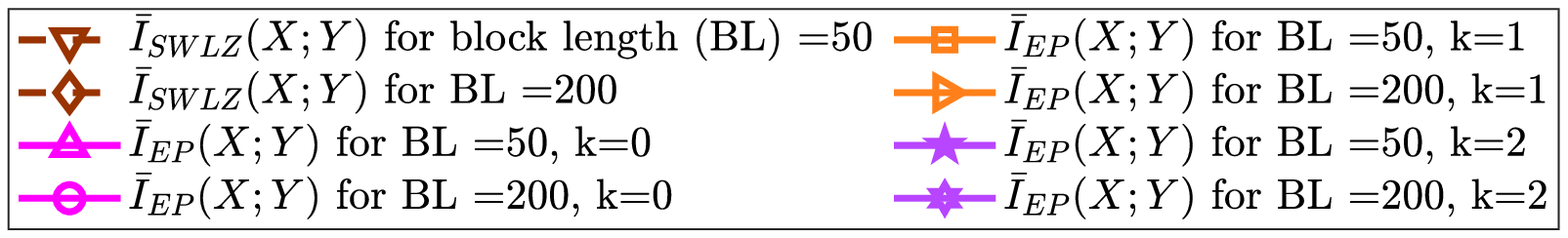}
}\vspace{-1.2em}\\
\centering
\subfigure{
  \includegraphics[width=0.6\columnwidth]{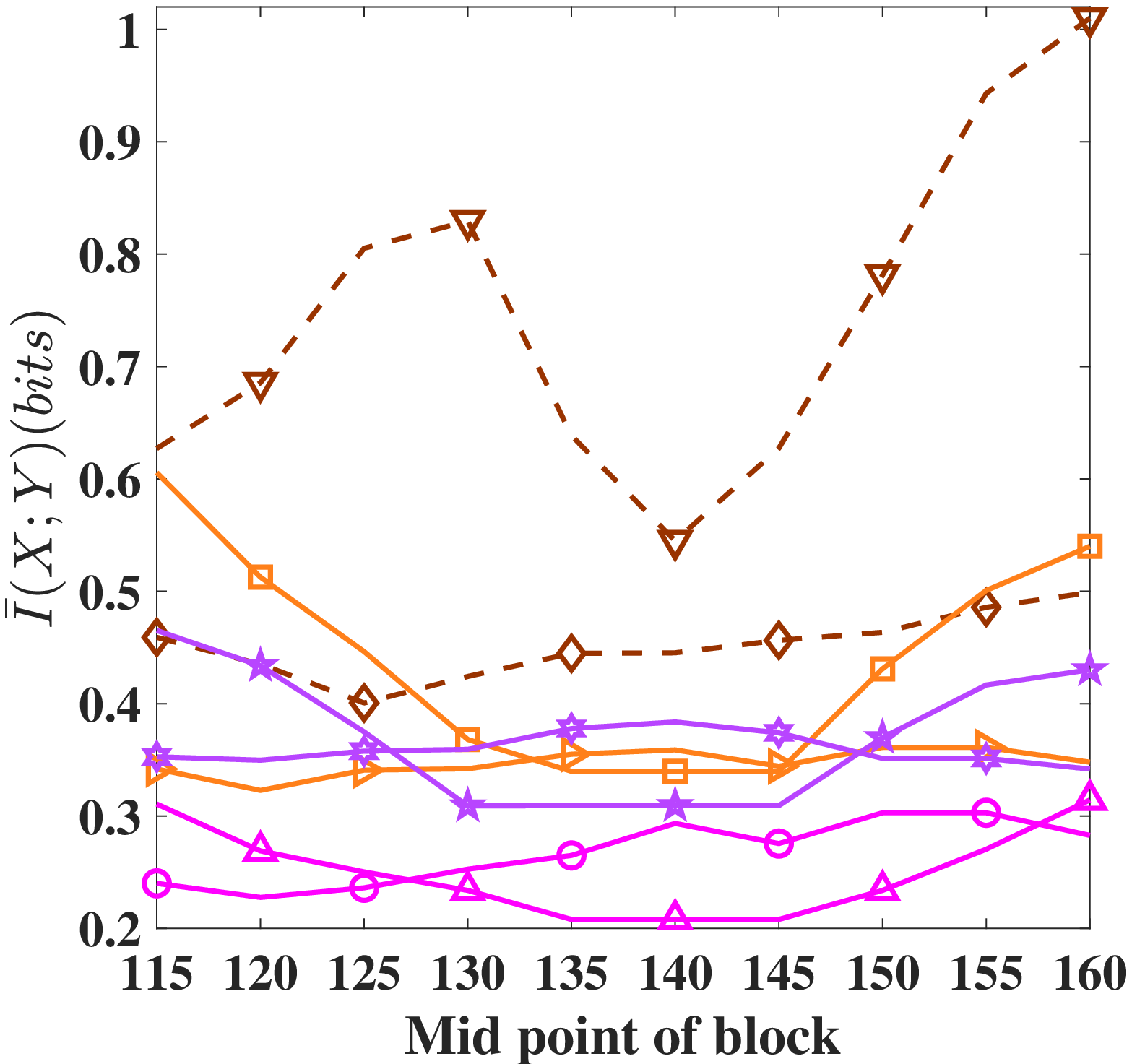}} \vspace{-1em}
 \caption{Variation in the proposed mutual information rate estimate $\bar{I}_{EP}(X;Y)$ and that in the competing estimate $\bar{I}_{SWLZ}(X;Y)$ with context lengths $k= 0,1,2$ against different window lengths considering neurons indexed 1, 9.}
\label{fig:estimates_windows_k012}
\end{figure}


%% file: figures/section3/f-rasterplot-neurons.tex
\begin{figure}[t!]
\centering
\includegraphics[width=\columnwidth]{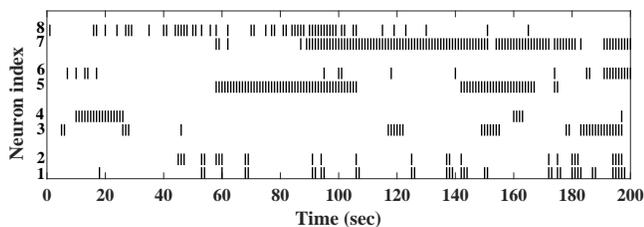}
 \vspace{-3.6em}
\caption{Raster plots of spike trains for neurons with indices 1-8 (see  Fig.\ref{fig:fluorescence-2}).}
\label{fig:Rasterplot_neurons}
\end{figure}

%% file: figures/section3/f07-temporal-variation-estimates.tex
\begin{figure*}[t!]
\centering
\begin{tabular}{cc}
\includegraphics[width=0.9\textwidth]{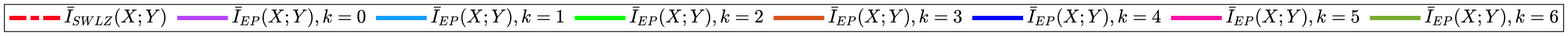}\vspace{-0.5em}\\
\subfigure[]{
\includegraphics[width=4.2cm]{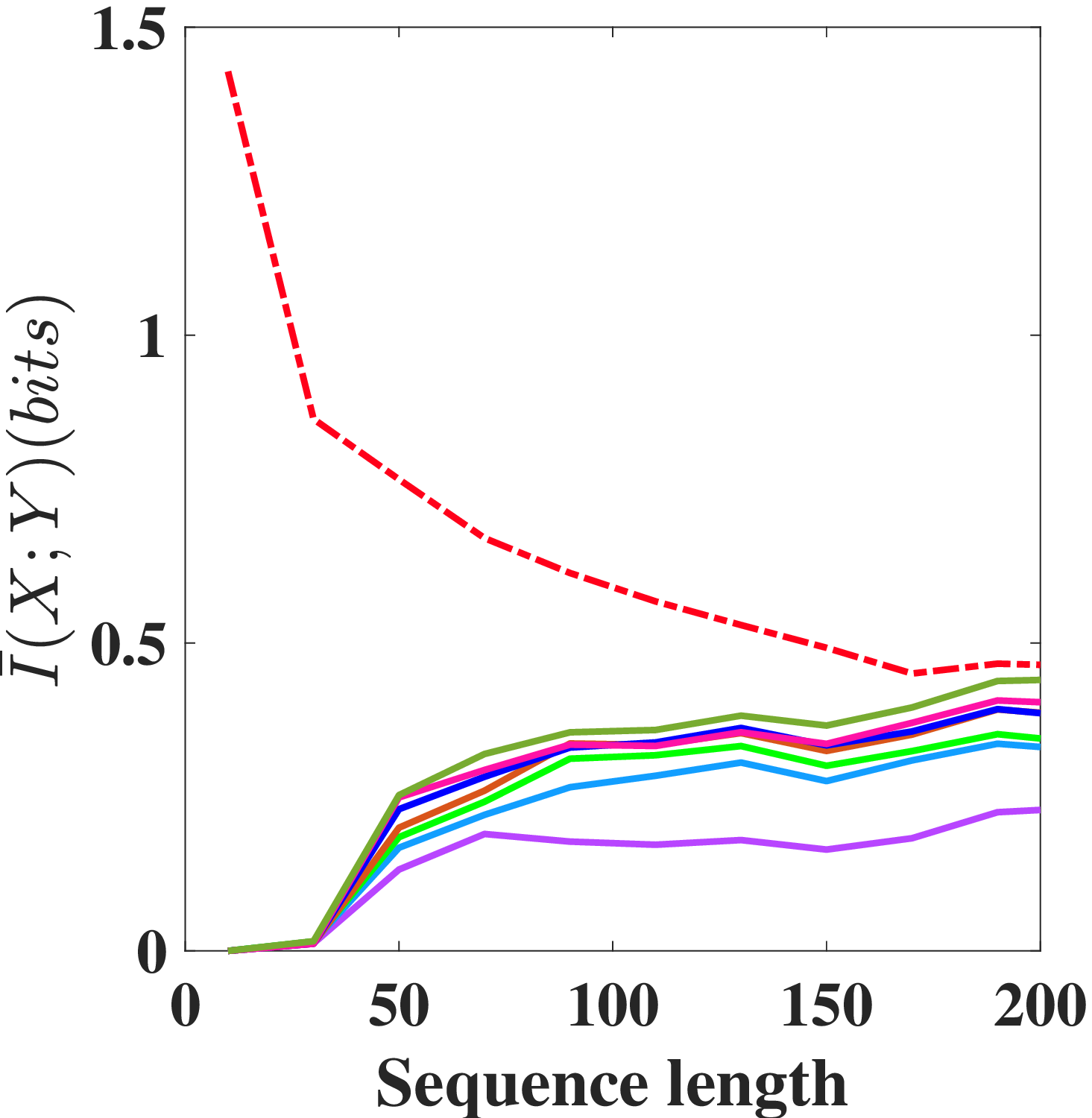}
\label{fig:nd1}\hspace{-1em}}
\subfigure[]{
\includegraphics[width=4.2cm]{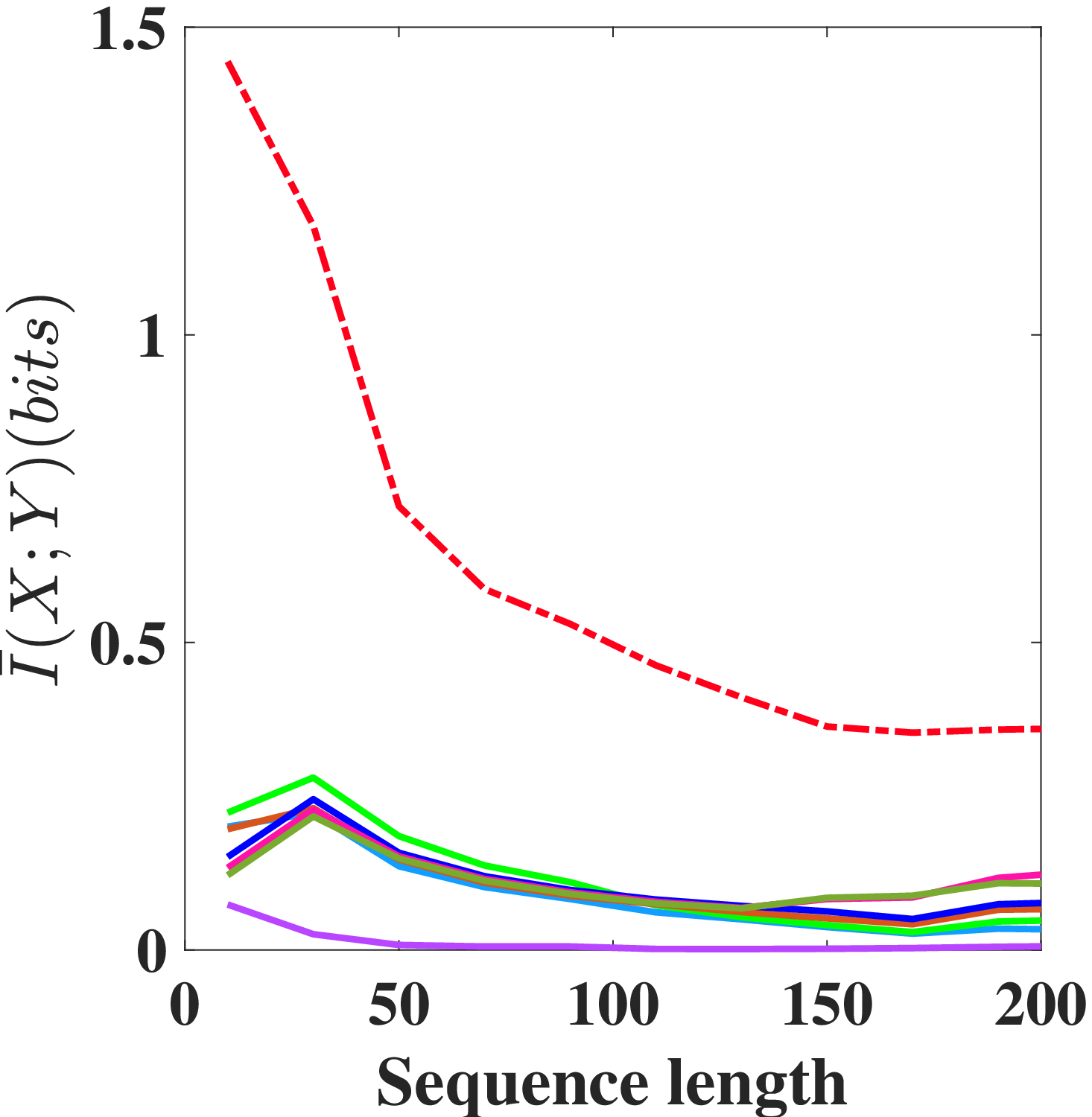}
\label{fig:nd2}\hspace{-1em}} 
\subfigure[]{
\includegraphics[width=4.2cm]{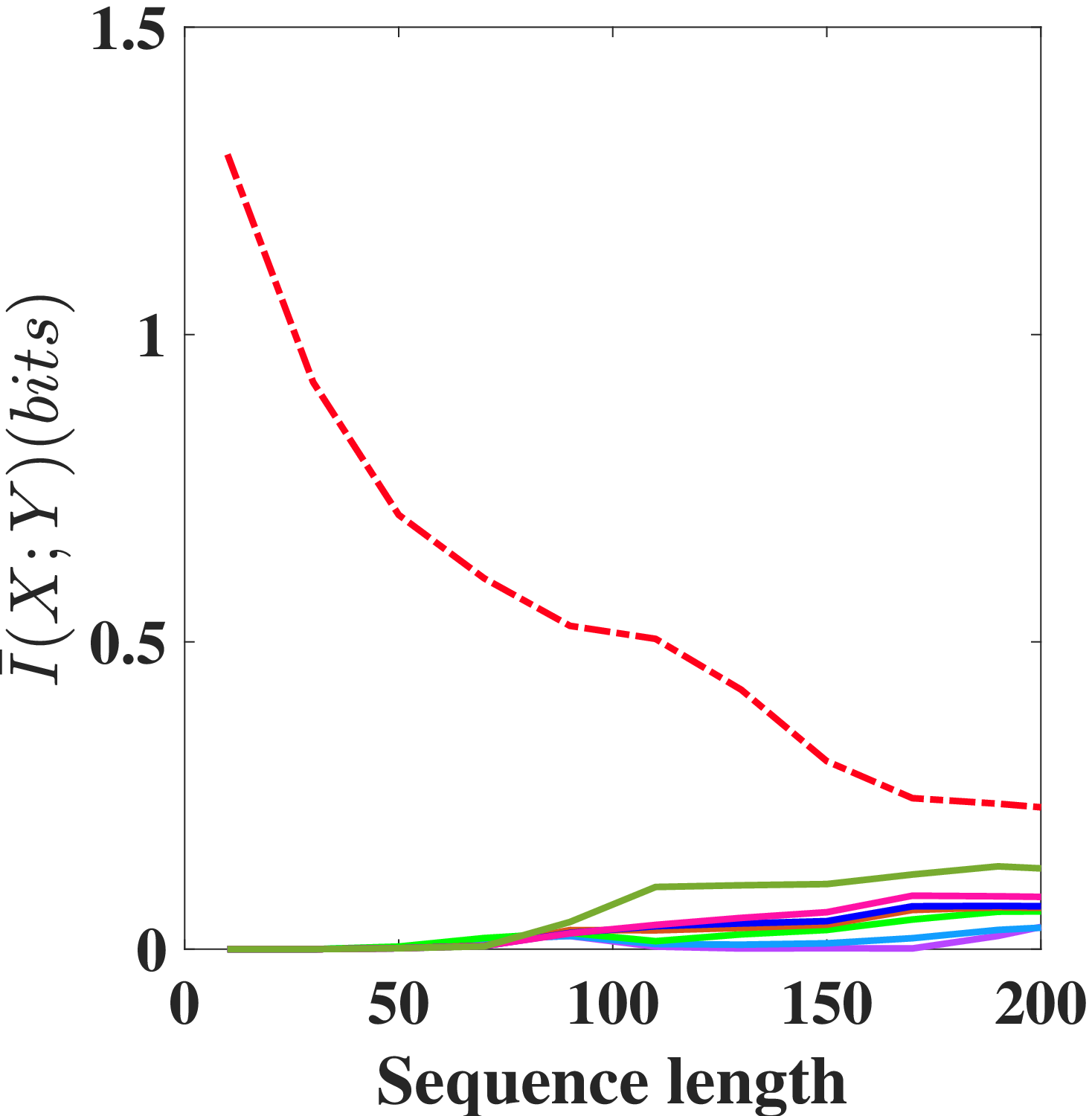}
\label{fig:nd3}\hspace{-1em}} 
\subfigure[]{
\includegraphics[width=4.2cm]{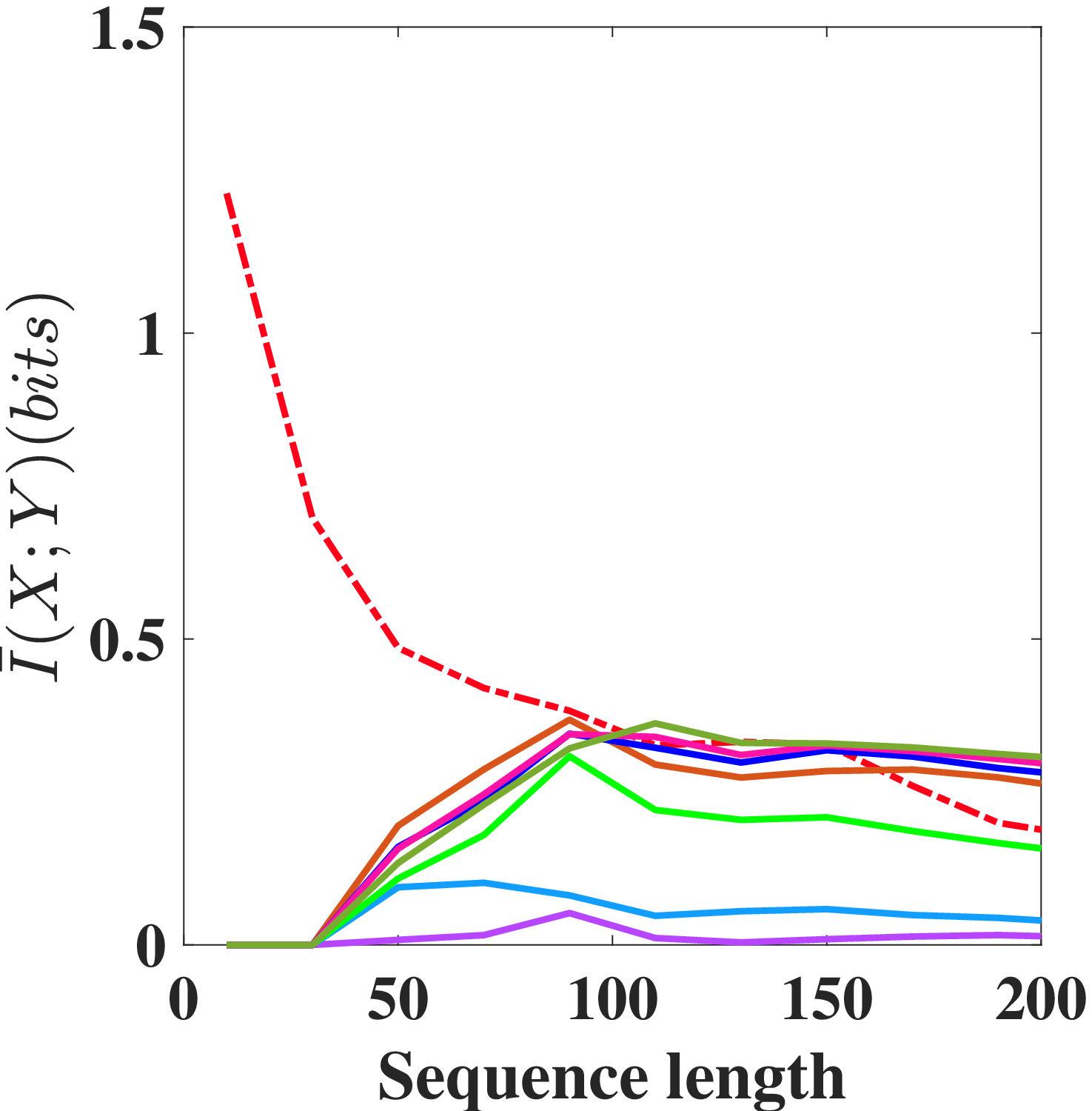}
\label{fig:nd4}}
\end{tabular} \vspace{-0.5em}
\caption{Temporal variation in mean of mutual information rate estimates based on SWLZ algorithm and empirical methods for $k=0-6$: (a),(b),(c),(d) for neuron pairs 1-2, 3-4, 5-6, 7-8, respectively.}
\label{fig:TempVariation}
\end{figure*}

%% file: section/conclusion.tex
\section{CONCLUSION}
\label{sec:conclusion}
In this paper, we proposed a fast yet accurate empirical method for estimating mutual information rate between calcium spike trains. Our method outperformed the existing LZ algorithms for lower order memory structures in synthetic as well as experimental spike trains. Furthermore, the heterogeneity in mutual information estimates was observed clearly in neurons. Even though, our method identified the time window required for efficient estimation and tracked temporal variation in information estimates, there is always scope for improvement. Especially, we need to do thorough analysis of complex process memory structures in large neuron populations. The current mutual information rate analysis could be helpful for understanding neuronal information encoding under normal and diseased conditions, and identifying disease signatures.

%% file: section/Iproof.tex
For the processs ($X(m),Y(m)$), we now prove (\ref{Eq:MIuv}) i.e., the mutual information rate $\bar{I}(X;Y) = I(u;v)$ irrespective of the choice of $m,n$.
 From (\ref{eq:muDef}), we have 
\begin{equation}
\begin{aligned}
\bar{I}(X;Y) = \frac{1}{n}H(X_1^n)+ \frac{1}{n}H(Y_1^n)-\frac{1}{n}H(X_1^n ,Y_1^n)\\
\end{aligned}
\label{eq:MarkovMI} 
\end{equation}
Let us elaborate each term as follows. 
\begin{equation}
\begin{aligned}
 \frac{1}{n}H(X_1^n) = \frac{1}{n} ( H(X_1) + \sum_{i=2}^n H(X_i | X_{i-m}^{i-1}))  \quad  \text{(by chain rule)} 
\end{aligned}
\label{eq:marginalx}
\end{equation}
 From (\ref{eq:markov1}), $H(X_1) = H(X_i|X_{i-m}^{i-1})=H(u)$  $\forall i, m$.\\
Hence (\ref{eq:marginalx}) becomes
\begin{equation}
\begin{aligned}
 \frac{1}{n}H(X_1^n)=  \frac{1}{n} ( H(u) + \sum_{i=2}^n H(u)) = H(u)
\end{aligned}
\label{eq:Hu}
\end{equation}
Similarly, \\
\begin{equation}
\begin{aligned}
\frac{1}{n}H(Y_1^n)=  \frac{1}{n} ( H(v) + \sum_{i=2}^n H(v)) = H(v)
\end{aligned}
\label{eq:Hv}
\end{equation}
Hence, the marginal entropy rates of $X(m)$, $Y(m)$ are equal to entropy $H(u)$, $H(v)$ of binary sources u, v respectively. $H(u) = H(v)= -0.5 \log_2 0.5-0.5 \log_2 0.5 = 1$ bit.
\begin{equation}
\begin{aligned}
\frac{1}{n}H(X_1^n,Y_1^n) = \frac{1}{n} ( H(X_1,Y_1) + \sum_{i=2}^n H(X_i,Y_i | X_{i-m}^{i-1},Y_{i-m}^{i-1})  
\end{aligned}
\label{eq:HXY}
\end{equation}
where
\begin{equation}
\begin{aligned}
H(X_i,Y_i|X_{i-m}^{i-1},Y_{i-m}^{i-1}) &= H(X_i|X_{i-m}^{i-1},Y_{i-m}^{i-1}) + H(Y_i | X_{i-m}^{i-1},Y_{i-m}^{i-1},X_i) \\
&=H(X_i|X_{i-m}^{i-1},Y_{i-m}^{i-1}) + H(Y_i | X_{i-m}^{i},Y_1^{i-1})\\
&=H(X_i|X_{i-m}^{i-1},Y_{i-m}^{i-1}) + H(Y_i | X_{i-m}^{i-1},u_i,Y_{i-m}^{i-1})\\
&=H(X_i|X_{i-m}^{i-1}) + H(v_i | u_i)\\
&=H(u_i) + H(v_i | u_i) =H(u_i,v_i) = H(u,v) \forall i,m
\end{aligned}
\label{eq:Hc}
\end{equation}
Similarly,
\begin{equation}
\begin{aligned}
H(X_1,Y_1)= H(u,v)
\end{aligned}
\label{eq:Hd}
\end{equation}
Substitute (\ref{eq:Hc}), (\ref{eq:Hd}) in (\ref{eq:HXY}) yields
\begin{equation}
\begin{aligned}
 \frac{1}{n}H(X_1^n,Y_1^n)= H(u,v)   
\end{aligned}
\label{eq:Huv}
\end{equation}
Hence, the joint entropy rate between sequences $X_1^n$, $Y_1^n$ generated from processes $X(m)$, $Y(m)$ respectively is equal to joint entropy $H(u,v)$ between the binary sources $u$, $v$.\\
Substitute (\ref{eq:Hu}), (\ref{eq:Hv}), (\ref{eq:Huv}) in (\ref{eq:MarkovMI}) yields
\begin{equation}
\begin{aligned}
\bar{I}(X;Y) &= H(u) + H(v)-H(u,v)= I(u;v) \quad \forall m
\end{aligned}
\label{Eq:Iuv}
\end{equation}
Hence, the mutual information rate $\bar{I}(X;Y)$ between $X(m)$, $Y(m)$ is equal to  mutual information ${I}(u;v)$ between binary sources $u$ and $v$.
 \begin{equation}
\begin{aligned}
H(u,v) &= -\sum_{u_i,v_i} pr(u_i,v_i) \log_2 pr(u_i,v_i) \\
&= -2(a_1) \log_2 (a_1) -2(a_2) \log_2 (a_2) \quad \quad \text{(from (\ref{EqUV}))} 
\end{aligned}
\label{Eq:ab}
\end{equation}
where $a_1 = 0.25 - p$ and  $a_1 = 0.25 + p$. \\
Substitute $H(u) = H(v)= 1$ bit and (\ref{Eq:ab}) in (\ref{Eq:Iuv}) yields
\begin{equation}
\begin{aligned}
\bar{I}(X;Y) = 2(1+(a_1) \log_2 (a_1) +(a_2) \log_2 (a_2) ) \quad \forall m
\end{aligned}
\label{Eq:IMarkov}
\end{equation}
This completes the proof.